\documentclass[prb,twocolumn,groupedaddress]{revtex4b4}
\usepackage{graphicx}

\begin{document}
\title{Saddle Points and Dynamics of Lennard-Jones Clusters, Solids and Supercooled Liquids}
\author{Jonathan P.~K.~Doye and David J.~Wales} 
\affiliation{University Chemical Laboratory, Lensfield Road, 
  Cambridge CB2 1EW, United Kingdom}
\date{\today}
\begin{abstract}
The properties of higher-index saddle points have been invoked in recent theories of 
the dynamics of supercooled liquids.
Here we examine in detail a mapping of configurations to saddle points using minimization of $|\nabla E|^2$,
which has been used in previous work to support these theories.
The examples we consider are a two-dimensional model energy surface and
binary Lennard-Jones liquids and solids.
A shortcoming of the mapping is its failure to divide the potential energy surface into basins of attraction 
surrounding saddle points, because there are many minima of $|\nabla E|^2$ that do not correspond
to stationary points of the potential energy. In fact, most liquid configurations are mapped 
to such points for the system we consider. 
We therefore develop an alternative route to investigate higher-index saddle points and
obtain near complete distributions of saddles for small Lennard-Jones clusters. 
The distribution of the number of stationary points as a function of the index is found to be 
Gaussian, and the average energy increases linearly with saddle point index
in agreement with previous results for bulk systems.
\end{abstract}
\maketitle

\section{Introduction}

The properties of the potential energy surface (PES), or landscape, of supercooled liquids
and glasses have been the focus of much recent theoretical research into glasses.
The origins of this approach date back to Goldstein, who suggested that
the dynamics could be separated into vibrational motion about a minimum on 
the PES and transitions between minima.\cite{Goldstein69} 
This idea led to the pioneering `inherent structure' approach of Stillinger and 
coworkers.\cite{StillW82,StillW84a}
In this approach the PES is partitioned into basins of attraction surrounding the minima
(inherent structures), where a basin of attraction is defined as the set of
points for which steepest-descent pathways lead to the same minimum.
This mapping allows a conceptual framework to be built in which the role of the minima can
be separated from the effects of vibrational motion.
The dynamics can then be viewed in terms of the transitions between these basins 
which occur when the system passes along a transition state valley.
Therefore, from the inherent structure point of view the 
key points on the PES are the minima and transition states, which are
defined as stationary points with Hessian index one, i.e.\ one negative eigenvalue.
Higher-index saddle points (with $I\geq 2$ negative eigenvalues) need not be considered 
in this description, because, by the Murrell-Laidler theorem, if two minima 
are connected by an index two saddle, then there must be a lower-energy path between 
them involving only true transition states.\cite{Murrell68}

This inherent structure approach has provided important insights into the behaviour of 
supercooled liquids and glasses,\cite{Still95,Debenedetti01} as well as clusters\cite{WalesDMMW00} and 
biomolecules.\cite{Brooks01}
For example, changes in the dynamics of supercooled liquids as the temperature is decreased
must correspond to descent down the PES to lower-energy minima.\cite{Sastry98} 
Furthermore, a careful investigation of the density dependence of the properties of the sampled minima 
has suggested how changes in the PES can lead to a change in the dynamics from strong to fragile.\cite{Sastry01}
This approach has also been applied to investigate non-equilibrium properties:
ageing involves a slow decrease in the energy of the sampled minima as 
the system heads towards equilibrium.\cite{Kob00,Sciortino01}

Much of the above work has focussed on the potential energy landscape as sampled under
particular conditions of density and temperature.
It is also of interest to determine the fundamental characteristics of the landscape,
which is, of course, independent of temperature, atomic masses and coordinate system.
For example, one of the most important properties is the distribution of minima. 
Exhaustive enumeration of the minima of small systems\cite{HoareM83,Tsai93a} seems
to confirm the theoretical conjecture that the number of minima increases 
exponentially with size.\cite{StillW82,Still99}
The distribution of minima as a function of the potential energy can also be obtained by inverting
simulation data. This inversion was first performed for a medium-sized cluster,\cite{Doye95a}
and later for model glasses\cite{Sciortino99a,Buchner99a} revealing that the 
distribution is Gaussian.
This technique has since been applied to supercooled water\cite{Scala00} and silica.\cite{SaikaVoivod01}
More recently, attention has begun to focus on the harder task of characterizing
the distributions of transition states and the resulting barriers.\cite{Middleton01,Mousseau00}

Alternatives to the inherent structure approach have been proposed. 
In the instantaneous normal mode theory, developed by Keyes and coworkers,\cite{Keyes97} 
the focus is on the spectrum of Hessian eigenvalues for instantaneous configurations.
It is argued that diffusion and the associated barriers are related to the negative eigenvalues 
of this spectrum, although this idea has been the subject of some debate.\cite{Gezelter97,Keyes98,Gezelter98}
Many of the negative eigenvalues result from the anharmonicity within a well,
however once these are removed, simulations have indicated that the number of
diffusive directions in a supercooled liquid tends towards zero near to the
mode-coupling temperature.\cite{Donati00,LaNave01}

Another recent proposal emphasizes the role of higher-index stationary points
on the PES, and attempts to explain the origins of strong and fragile liquids in 
these terms.\cite{Cavagna01} 
One of the underlying ideas is that as the size of the system becomes large, most 
of the configuration space volume in a basin of attraction is concentrated near to 
the borders of that basin, and so a randomly chosen point is more likely to 
lie closer to a saddle point than to a minimum.\cite{Kurchan96} Therefore, the proposal 
is to divide the potential energy surface into ``basins of attraction'' 
that surround stationary points of any index.
However, with the steepest-descent mapping of the inherent structure approach, the
basins of attraction only surround minima, and convergence to a higher-index saddle point
can only occur when the starting point lies exactly on the boundary between two basins 
of attraction. The volume associated with these boundaries is of measure zero.
Therefore, a different mapping is required to divide configuration space in the desired way.
Borrowing a trick 
that has been previously used to locate transition states,\cite{McIver72,Weber85}
a mapping has been suggested in which steepest-descent paths
on the function $|\nabla E|^2$, the modulus of the gradient of the energy squared, 
are followed.\cite{Angelani00,Broderix00}
Stationary points on the PES of any index correspond to minima of this new function.

From the above mapping it was suggested that below the mode-coupling\cite{Gotze91,Kob97,Gotze99} 
temperature, $T_c$, the system samples minima but that above this temperature the 
average saddle point index increases linearly with temperature.\cite{Angelani00,Broderix00} 
Such behaviour has been interpreted as a transition from dynamics between basins of minima to 
dynamics between basins of saddles. This approach is becoming more widely 
applied.\cite{DiLeonardo01,Angelani01,Scala01,Grigera01,Shah01} For example, it has been used to
analyse the dynamics of ageing. After a quench or crunch (a sudden increase in density)
to a state that lies below $T_c$ for that density, 
initially the system is associated with saddle points whose index decreases 
logarithmically with time until a crossover time is reached when the system resides near to 
minima.\cite{DiLeonardo01,Angelani01}

Here we look in more detail at the $|\nabla E|^2$ mapping and how well it achieves its aim
of dividing configuration space into neighbourhoods around stationary points of any index.
We first examine in Section \ref{sect:MB} a model two-dimensional energy surface
that can easily be visualized. Then, in Section \ref{sect:BLJ} we study the properties of the
mapping for a much-studied binary Lennard-Jones system.
Given the problems with the $|\nabla E|^2$ mapping we then follow an alternative approach
to studying the properties of higher-index saddle points. In section \ref{sect:clusters} 
we obtain near complete distributions of saddle points for small Lennard-Jones clusters
and then analyse their properties. Finally, we conclude with a discussion of some of the
issues raised by our results in relation to recent work.

\section{M\"{u}ller-Brown surface}
\label{sect:MB}

We first examine the effect of the $|\nabla E|^2$ mapping for a model two-dimensional
energy surface that we can easily visualize. 
We use the M\"{u}ller-Brown surface,\cite{MullerBrown} which has the form:
\begin{equation}
E(x,y)=\sum_{i=1}^4 \exp[a_i(x-x_i^0)^2+b_i(x-x_i^0)(y-y_i^0)+c_i(y-y_i^0)^2],
\end{equation}
where
\begin{eqnarray}
&A=(-200,-100,-170,15)\quad &a=(-1,-1,-6.5,0.7) \nonumber \\
&b=(0,0,11,0.6)\quad &c=(-10,-10,-6.5,0.7)\nonumber  \\
&x^0=(1,0,-0.5,-1)\quad &y^0=(0,0.5,1.5,1).
\end{eqnarray}
This surface has been used as a test system for local optimization algorithms
and its properties have been thoroughly examined.\cite{Sun93c,Sun93d,Ruedenberg94,Wales94b,Passerone01}

\begin{figure}
\begin{center}
\includegraphics[width=6.2cm]{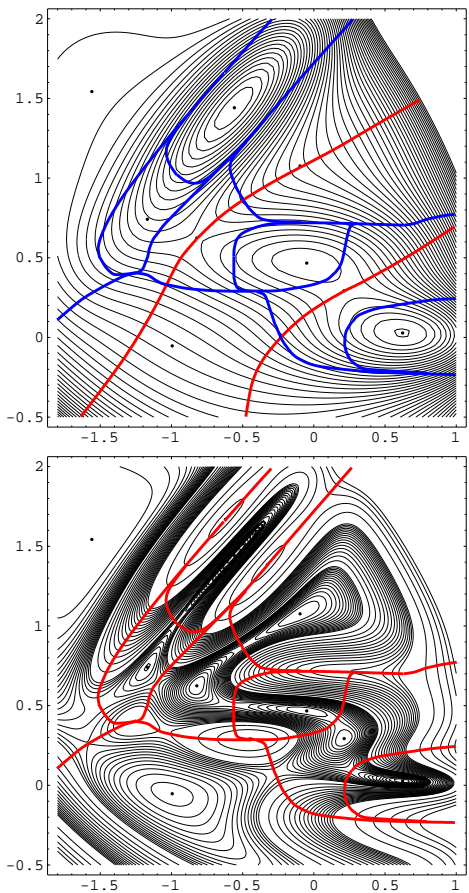}
\caption{\label{fig:MB} Contour diagrams of (a) $E$ and (b) $|\nabla E|^2$ 
for the M\"{u}ller-Brown surface.  The red lines divide the surfaces into basins of 
attraction surrounding each minimum. In (a) the basin boundaries 
of $|\nabla E|^2$ have been superimposed in blue. 
Points have been added corresponding to in (a) the minima of $|\nabla E|^2$ and 
in (b) the maxima and minima of $|\nabla E|^2$. In (b) contours occur
every 1000 in the range 0 to 20000, then every 5000 up to 100000 and
finally every 50000 beyond that range.}
\end{center}
\end{figure}

Contour plots of $E$ and $|\nabla E|^2$ are shown in Figure \ref{fig:MB} and information concerning 
the minima of $|\nabla E|^2$ are given in Table \ref{table:MB}.\cite{DJWwhoops} 
The section of the energy surface that we consider has five stationary points: 
three minima and the two transition states that connect them.
Stationary points of the $|\nabla E|^2$ surface satisfy $2{\bf H}\,{\bf g}={\bf 0}$, where ${\bf H}$ is the Hessian
(second derivative matrix) and ${\bf g}=\nabla E$ is the gradient vector. 
Obviously, stationary points of the PES
have $|\nabla E|^2=0$, and so correspond to minima of $|\nabla E|^2$. 
However, there are additional minima on the $|\nabla E|^2$ surface with $|\nabla E|^2>0$,
where ${\bf g}$ is an eigenvector of ${\bf H}$ with zero eigenvalue, 
i.e.\  there is an inflection point in the direction of the gradient. 
We will refer to these two types of $|\nabla E|^2$ minima by the labels SP 
(stationary point of $E$) and NSP (non-stationary point of $E$).
There are four such NSP's on the M\"{u}ller-Brown surface.
The possibility of NSP's has been previously noted in 
Refs.\ \onlinecite{Cavagna01}, \onlinecite{Angelani00} and \onlinecite{Broderix00},
however it was claimed that their effect was negligible.\cite{Angelani00} 

A further property of the NSP's is that they must lie on a gradient extremal, 
one definition of which is ${\bf H}\,{\bf g}=\lambda\,{\bf g}$. Gradient extremals are curves 
for which each point is an extremum of $|\nabla E| \equiv |{\bf g}|$
along the corresponding energy contour. 
The gradient extremals for the M\"{u}ller-Brown surface
have been calculated in Ref.\ \onlinecite{Sun93d}.

The basins of attraction associated with the SP's and NSP's on the 
$|\nabla E|^2$ surface are shown in Figure \ref{fig:MB}. 
For the region of configuration space that we depict here, the majority of this space 
corresponds to basins of attraction associated with NSP's. 
Although much of this configuration space has a relatively high energy, 
the NSP basins of attraction do extend into some low energy regions. 
In particular, NSP1 is lower in energy than the transition state SP5, and on moving
away from the minimum SP1 along the softest mode, the first new basin of attraction that is 
encountered corresponds to NSP1. Furthermore, the basins of attraction 
associated with NSP2 and NSP3 extend below the energy of SP5 into regions corresponding
to the walls of the basin of attraction associated with SP1 on the original surface.
Minimization of $|\nabla E|^2$ for points from these regions will lead to a considerable 
increase in energy.

It is clear from Figure \ref{fig:MB} that the $|\nabla E|^2$ surface is 
much more rugged than the original energy surface.
Firstly, the surface has more minima.
Secondly, the ratio of maximum to minimum non-zero Hessian eigenvalues of the $|\nabla E|^2$ function 
at SP's has a magnitude that is roughly the square of the corresponding ratio for stationary points of $E$,
because the second derivative of $|\nabla E|^2$ includes a product of the original Hessian matrix. 
For SP1 this ratio is $\sim$100 and so the well surrounding SP1 on the $|\nabla E|^2$ surface is 
extremely asymmetric and narrow.

This second feature of the $|\nabla E|^2$ surface has 
important practical consequences.
In the language of optimization theory, such points are said to be ill-conditioned,\cite{Acton} 
and so even simple minimization of $|\nabla E|^2$ is likely to be rather slow. These effects will be
further examined when we consider a binary LJ system in Section \ref{sect:BLJ}.

Although the M\"{u}ller-Brown surface is not a physical example, it does suggest
that the division of configuration space into basins of attraction surrounding
the minima of $|\nabla E|^2$ could be problematic.

\begin{table}
\caption{\label{table:MB} Minima of $|\nabla E|^2$ for the Muller-Brown surface. Those that are also
stationary points of $E$ are labelled SP, and those that are not are labelled NSP. 
The index, $I$, is the number of negative eigenvalues of the Hessian at that point.}
\begin{ruledtabular}
\begin{tabular}{cccccc}
 &  I & $E$ & $|\nabla E|^2$ & $x$ & $y$ \\
\hline
 SP1 &  0 & $-146.700$ &     0.0 & $-0.558$ &  1.442 \\
 SP2 &  0 & $-108.167$ &     0.0 &  0.623 &  0.028 \\
 SP3 &  0 &  $-80.768$ &     0.0 & $-0.050$ &  0.467 \\
 SP4 &  1 &  $-72.249$ &     0.0 &  0.212 &  0.293 \\
 SP5 &  1 &  $-40.665$ &     0.0 & $-0.822$ &  0.624 \\
 NSP1 & 0 &  $-56.235$ & 10892.8 & $-1.169$ &  0.741 \\
 NSP2 & 0 &   19.057 &   175.2 & $-1.559$ &  1.543 \\
 NSP3 & 1 &   21.394 &  5018.2 & $-0.097$ &  1.076 \\
 NSP4 & 0 &   27.070 &  6533.2 & $-0.995$ & $-0.053$ \\
\end{tabular}
\end{ruledtabular}
\end{table}

\section{Bulk Binary Lennard-Jones}
\label{sect:BLJ}

Binary Lennard-Jones (BLJ) mixtures have been extensively studied in an effort to elucidate the complex phenomenology of
glasses, as they do not crystallize on the molecular dynamics (MD) time
scale when suitably 
parameterized.\cite{Jonsson,Kob94,Kob95,Kob95b,Sastry98,Buchner99a,Donati99,Angell99,Sciortino99a,Kob00,Schroder00,%
Donati00,Mousseau00,Sastry00a,Sastry00b,Sciortino00,Angelani00,Broderix00,Buchner00,Middleton01,Sciortino01,Sastry01}
Most of these investigations have reported changes in behaviour around the 
critical temperature predicted by mode-coupling theory,\cite{Gotze91,Kob97,Gotze99}
which is calculated as $T_c\approx0.435$ for the mixture we consider here.\cite{Kob94,Kob95}
For example, Sastry, Debenedetti and Stillinger\cite{Sastry98} presented evidence
that non-exponential relaxation starts below about $T=1$,
while the height of the effective barriers to relaxation increases significantly
around $T=0.45$. They view the regions below $T=1$ and
$T=0.45$ as `landscape-influenced' and `landscape-dominated' regimes, respectively.
They also found that local minima
obtained by quenching from configurations generated at $T=0.5$ could escape to different local minima much more easily
than local minima obtained from
configurations generated at $T=0.4$.

Using an instantaneous normal modes picture Donati, Sciortino and Tartaglia also 
concluded that activated dynamics becomes important around $T_c$.\cite{Donati00}
Schr\o der {\it et al.}~found that the liquid dynamics could be separated into transitions between minima 
and vibrational motion\cite{Goldstein69} at a similar temperature,\cite{Schroder00} and correlated
motions of atoms in groups that grew with decreasing temperature were
reported by some of the same workers.\cite{Donati99}
Two different groups have recently reported that the typical Hessian index of stationary
points sampled by BLJ systems extrapolates to zero, again around $T_c$,\cite{Angelani00,Broderix00}
and we will present a more detailed investigation of this behaviour below.
Our results show that transition states are still accessible below $T_c$, but support the
general consensus that the PES sampled by the BLJ system
changes in character somewhere around $T_c$.

The BLJ mixture that we consider involves
a 256-atom supercell containing 205 (80\%) A atoms, and 51 (20\%)
B atoms, with parameters $\sigma_{AA}=1.0$, $\sigma_{AB}=0.8$,
$\sigma_{BB}=0.88$, $\epsilon_{AA}=1.00$, $\epsilon_{AB}=1.5$, and
$\epsilon_{BB}=0.5$.\cite{Kob94}  The units of distance and energy were taken as
$\sigma_{AA}$ and $\epsilon_{AA}$. The chosen box length gives a fixed
number density of $1.2\,\sigma_{AA}^{-3}$, and a
cutoff of $2.5\,\sigma_{AA}$ was used along with the minimum image convention 
and a shifting/truncation scheme that ensures continuity of the energy and its first derivative, 
as in previous work.\cite{Sastry98,Stoddard73}

Standard microcanonical molecular dynamics (MD) simulations of $10^5$ equilibration steps, followed by
$10^6$ data collection steps, were first run using the Verlet propagator
at a series of increasing total energies (Table \ref{tab:MD}). 
The starting point for the first run was the lowest-energy minimum found in previous work,\cite{Middleton01}
and subsequent runs used the final configuration of the previous trajectory as the starting point.
A time step of $0.003$ reduced units was employed in each case. Every 1000th configuration from the data
collection phase was saved and used as a starting point for the following geometry optimizations:
(1) minimization using a modified version of Nocedal's LBFGS algorithm,\cite{Liu89}
(2) a transition state search using hybrid eigenvector-following,\cite{Munro99,WalesDMMW00,Kumeda01}
(3) minimization of $|\nabla E|^2$ using Nocedal's LBFGS algorithm.\cite{Liu89}
The first two searches on the conventional PES employ standard techniques\cite{WalesDMMW00} and were followed
by between one to three full eigenvector-following steps to converge the root-mean-square (RMS) force
below $10^{-7}$ reduced units and check the Hessian index of the stationary point, defined as
the number of negative Hessian eigenvalues. All the searches in (1) and (2) are tightly converged
to stationary points of the required index with the above tolerance, which is more
stringent than the criteria used in Ref.\ \onlinecite{Shah01}.

As pointed out in Section \ref{sect:MB} the ratio of the maximum to the minimum eigenvalue 
of $|\nabla E|^2$ near to an SP makes minimization of $|\nabla E|^2$ laborious.
There is a further problem because the second derivatives of the
shifted-truncated BLJ potential\cite{Stoddard73} are discontinuous at the cut-off, and hence the derivatives of $|\nabla E|^2$
have corresponding discontinuities. Nevertheless, it is possible to converge the $|\nabla E|^2$ minimizations
to an root-mean-square force below $10^{-5}$, at which point the value of $|\nabla E|^2$ is generally converged to 
at least nine decimal places. Such accuracy should be acceptable for the present purposes, and was achieved
by fixing the neighbour list during minimization
for SP's and NSP's close to convergence but suffering from discontinuities.
Typically, these minimizations of $|\nabla E|^2$ involve two orders of magnitude more
steps than minimizations on the original PES.

None of the previous work reporting results of $|\nabla E|^2$ minimizations\cite{Angelani00,Broderix00} has 
specified any details of the calculations, such as the algorithms employed, the convergence criteria
or the number of saddles actually found. 
We are therefore unable to provide any detailed comparisons.

\begin{table*}
\begin{center}
\caption{\label{tab:MD}
Mean total energy, $E$, potential energy, PE, kinetic energy, KE, and kinetic equipartition temperature, $T$, for 
the MD runs. The $\pm$ values represent one standard deviation. $\#{\rm min}$, $\#{\rm ts}$
and $\#{\rm G2}$ are the number of distinct minima,
transition states and stationary point of $|\nabla E|^2$ found for $10^3$ searches 
(excluding permutational isomers), as described in the text. 
\%SP and \%NSP are the percentage of quenches on the $|\nabla E|^2$ surface that converged to
stationary points and non-stationary points of $E$, respectively (out of $10^3$ total). }
\begin{ruledtabular}
\begin{tabular}{crllcccccc}
 run & $E$\hphantom{crapcr}  & \hphantom{crap}PE & \ \ KE & $T$ & $\#{\rm min}$ & $\#{\rm ts}$ & $\#{\rm G2}$ & \%SP & \%NSP \\
\hline
\noalign{\smallskip}
\noalign{\centerline{\bf 256-atom supercell}}
\noalign{\smallskip}
1 & $-1699.879\pm0.002$ & $-1753\pm2$ & \ $53\pm2$  & $0.138\pm0.005$ &  54 & 274  & $740$  &  $1.6$ & $98.4$ \\
2 & $-1599.268\pm0.005$ & $-1704\pm4$ &  $105\pm4$  & $0.276\pm0.010$ &  46 & 509  & $909$  &  $9.6$ & $90.4$ \\
3 & $-1500.070\pm0.008$ & $-1656\pm6$ &  $156\pm6$  & $0.409\pm0.015$ & 280 & 789  & $1000$ &  $1.8$ & $98.2$ \\
4 & $-1399.487\pm0.011$ & $-1594\pm7$ &  $194\pm7$  & $0.510\pm0.019$ & 987 & 1000 & $1000$ &  $3.2$ & $96.8$ \\
5 & $-1301.872\pm0.015$ & $-1540\pm9$ &  $238\pm9$  & $0.625\pm0.023$ & 984 & 1000 & $1000$ &  $3.0$ & $97.0$ \\
6 & $-1201.004\pm0.020$ & $-1485\pm11$ & $284\pm11$ & $0.745\pm0.028$ & 991 & 1000 & $1000$ &  $4.8$ & $95.2$ \\
7 & $-1098.883\pm0.025$ & $-1431\pm12$ & $332\pm12$ & $0.871\pm0.032$ & 995 & 1000 & $1000$ &  $5.9$ & $94.1$ \\
8 & $-999.982\pm0.031$  & $-1380\pm14$ & $380\pm14$ & $0.997\pm0.036$ & 994 & 1000 & $1000$ &  $4.8$ & $95.2$ \\
\noalign{\smallskip}
\noalign{\centerline{\bf 320-atom crystal}}
\noalign{\smallskip}
1 & $-2192.679\pm0.002$ & $-2270\pm2$  & $78\pm2$    & $0.162\pm0.002$ &  1   & 39  & $1$    & $100.0$ & $0.0$ \\
2 & $-1992.690\pm0.007$ & $-2175\pm6$  & $182\pm6$   & $0.382\pm0.012$ &  1   & 49  & $2$    & $99.4$  & $0.6$ \\
3 & $-1792.710\pm0.013$ & $-2082\pm9$  & $289\pm9$   & $0.606\pm0.019$ &  1   & 56  & $10$   & $91.0$  & $9.0$ \\
4 & $-1592.739\pm0.020$ & $-1989\pm12$ & $397\pm12$  & $0.831\pm0.026$ &  1   & 64  & $55$   & $66.4$  & $33.6$ \\
5 & $-1392.773\pm0.034$ & $-1889\pm16$ & $496\pm16$  & $1.041\pm0.034$ &  187 & 721 & $860$  & $25.5$  & $74.5$ \\
\end{tabular}
\end{ruledtabular}
\end{center}
\end{table*}

\begin{table*}
\begin{center}
\caption{\label{tab:diffs}
Mean energy differences, $\Delta$, and displacements, $D$, between the starting point and the converged geometry
after searching for minima (min), transition states (ts) and minimizing $|\nabla E|^2$ (G2). 
$i_{\rm SP}$ and $i_{\rm NSP}$ are the fractions of negative Hessian eigenvalues after
minimizing $|\nabla E|^2$, split into stationary points and non-stationary points of $E$, respectively.
The $\pm$ values represent one standard deviation. 
}
\begin{ruledtabular}
\begin{tabular}{ccrcrcrrr}
           run & $\Delta E_{\rm min}$ & $D_{\rm min}$\hphantom{crap} & $\Delta E_{\rm ts}$ & $D_{\rm ts}$\hphantom{crap} 
               & $\Delta E_{\rm G2}$ & $D_{\rm G2}$\hphantom{crap}  & 
               $i_{\rm NSP}\times10^3$\hphantom{} & $i_{\rm SP}\times10^3$\hphantom{} \\
\hline
\noalign{\smallskip}
\noalign{\centerline{\bf 256-atom supercell}}
\noalign{\smallskip}
1 &  $216\pm17$ & $22.0\pm1.6$ &  
     $208\pm17$ & $23.1\pm2.0$ & 
     $199\pm16$ & $24.2\pm1.5$ & $5.1\pm1.7$ & $4.2\pm2.1$ \\
2 &  $261\pm22$ & $25.2\pm0.8$ & 
     $253\pm22$ & $25.4\pm1.4$ &
     $245\pm22$ & $25.6\pm1.2$ & $5.4\pm2.0$ & $5.1\pm1.3$ \\
3 &  $303\pm64$ & $25.4\pm1.7$ & 
     $299\pm62$ & $25.7\pm1.6$ &
     $286\pm85$ & $26.1\pm1.7$ & $6.7\pm2.4$ & $5.8\pm1.7$ \\
4 &  $370\pm37$ & $81.3\pm6.2$ &
     $365\pm36$ & $81.3\pm6.1$ &
     $330\pm34$ & $81.3\pm6.2$ & $15.0\pm3.9$ & $12.7\pm5.5$ \\
5 &  $425\pm40$ & $144.5\pm10.2$ &
     $421\pm40$ & $144.5\pm10.2$ &
     $366\pm38$ & $144.5\pm10.1$ & $21.3\pm4.4$ & $19.6\pm5.2$\\
6 &  $485\pm46$ & $224.7\pm12.6$ &
     $480\pm46$ & $224.7\pm12.6$ &
     $403\pm44$ & $224.7\pm12.6$ & $28.0\pm5.1$ & $27.8\pm6.0$\\
7 &  $544\pm48$ & $320.3\pm16.9$ &
     $539\pm48$ & $320.3\pm17.0$ &
     $439\pm45$ & $320.3\pm16.9$ & $34.5\pm5.5$ & $34.4\pm7.1$\\
8 &  $602\pm56$ & $426.5\pm22.3$ &
     $598\pm56$ & $426.5\pm22.3$ &
     $475\pm52$ & $426.5\pm22.3$ & $41.0\pm5.6$ & $42.1\pm5.5$\\
\noalign{\smallskip}
\noalign{\centerline{\bf 320-atom crystal}}
\noalign{\smallskip}
1 &  $75\pm2$   & $8.8\pm0.04$ & 
     $53\pm8$   & $9.4\pm1.1$ &
     $75\pm2$   & $8.8\pm0.3$ & --- \ \ \ \ & $0.0\pm0.0$ \\
2 &  $170\pm6$  & $8.8\pm0.05$ & 
     $144\pm10$ & $9.7\pm1.0$ &
     $170\pm6$  & $8.8\pm0.05$ & $0.0\pm0.0$ & $0.0\pm0.0$ \\
3 &  $263\pm9$  & $8.9\pm0.1$ & 
     $240\pm13$ & $9.9\pm1.5$ &
     $263\pm9$  & $8.9\pm0.1$ & $0.0\pm0.2$ & $0.0\pm0.2$ \\
4 &  $356\pm12$ & $9.0\pm0.1$ & 
     $334\pm16$ & $9.8\pm1.3$ &
     $352\pm13$ & $9.0\pm0.2$ & $0.2\pm0.5$ & $0.0\pm0.2$ \\
5 &  $443\pm15$ & $15.4\pm3.7$ & 
     $432\pm18$ & $15.7\pm3.6$ &
     $425\pm20$ & $15.7\pm3.7$ & $1.6\pm1.5$ & $1.2\pm1.3$ \\
\end{tabular}
\end{ruledtabular}
\end{center}
\end{table*}

The glass transition temperature for this system under our simulation conditions
lies between $0.4$ and $0.5$, as is evident from the caloric curve 
(not illustrated) and jumps in various quantities tabulated 
in Tables \ref{tab:MD} and \ref{tab:diffs}. 
Of course, the precise temperature at which the glass transition occurs 
depends upon the rate at which the temperature is changed.
The number of different minima
and transition states located from the $10^3$ different starting points decreases markedly on glass formation. 
However, a significant number of distinct minima and transition states are 
sampled below the glass transition, as expected from 
previous work that revealed large numbers of low barrier ``non-diffusive'' pathways for this system.\cite{Middleton01}
The fraction of negative eigenvalues, $i=I/(3N-3)$, 
(the three zero eigenvalues corresponding to translations are excluded)
located by minimizing $|\nabla E|^2$ jumps by a factor of about two on melting, and continues 
to rise approximately linearly at higher temperature. 
This result is in line with previous calculations for supercooled
liquids,\cite{Angelani00,Broderix00,Grigera01} and also with studies 
based upon instantaneous normal mode analysis.\cite{Donati00}
However, in contrast to the claims of Refs.\ \onlinecite{Angelani00}, \onlinecite{Broderix00} and 
\onlinecite{Grigera01}, the value of $i$ does not vanish, even for our 
low-temperature glassy configurations. 

Throughout the temperature range studied here the vast majority of minimizations of 
$|\nabla E|^2$ converge to NSP's. 
The largest percentage of SP's occurs for run 2, but this is probably a fluctuation caused by the
non-ergodic nature of the low temperature simulations. Similarly, in Ref.\ \onlinecite{Broderix00} 
NSP's were said to be ``frequently sampled'' and these points were excluded from the calculated properties.
By contrast, in Ref.\ \onlinecite{Angelani00} the number of NSP's was said to be ``negligible with respect to the 
number of true saddles.'' 
However, stimulated by the current results, the authors of this paper have now identified the 
source of this discrepancy. A reanalysis of their original results has 
confirmed that the majority of the $|\nabla E|^2$ minima that they found were in fact NSP's.\cite{Scala01} 

For each run we find the mean energy differences between the
starting point and the structures obtained after searching for minima (min), 
transition states (ts) and minimizing $|\nabla E|^2$ (G2)
are in the order $\Delta E_{\rm min} > \Delta E_{\rm ts} > \Delta E_{\rm G2}$.
Hence, in terms of energy, the system is closer to points found by 
minimizing $|\nabla E|^2$ than it is to minima and transition states, 
especially at high temperature. This result is unsurprising.
By the same logic as the Murrell-Laidler theorem\cite{Murrell68}
one expects the average potential energy of a stationary point to increase with $I$.

However, the Euclidean distance between the above points is practically the same for all three searches 
(we checked that the centre-of-mass did not change during geometry optimization).
In terms of distance, the system seems to be equally close to a minimum, a true transition state, 
and an SP or NSP at all temperatures. 
Therefore, it cannot simply be claimed that the $|\nabla E|^2$ mapping 
takes the system to its nearest stationary point.
These results contrast with the behaviour for a spin glass, namely the p-spin spherical model,
for which the closest minimum is significantly further away from an 
equilibrium configuration than is the closest saddle when the system is above 
the glass transition.\cite{Cavagna01b} 
For this model the distance to the closest saddle was also found to be independent of temperature, 
again differing from our results for a structural glass (Table \ref {tab:diffs}).

To check that our results do not depend significantly on the minimization algorithm employed 
we repeated some of the $|\nabla E|^2$ minimizations using
a true steepest-descent approach. Both fifth-order Runga-Kutta and Bulirsch-Stoer algorithms were 
considered,\cite{Recipes} with the former method proving to be the more efficient, 
although it still required between 10$^2$ and $10^3$ times more steps than LBFGS minimization.\cite{Liu89} 
To reduce the computational cost we used
fifth-order Runga-Kutta integration of the steepest-descent equations for order $10^5$ steps and then switched
to LBFGS minimization.\cite{Liu89} For 100 regularly spaced starting points from trajectories 1 and 8 the
statistics produced by this alternative minimization scheme were very similar to the results in Tables
\ref{tab:MD} and \ref{tab:diffs}.

We have also generated results for the BLJ crystal with space group {\it I4/mmm},
which we have recently described elsewhere.\cite{Middleton01b} Here the supercell consists of
320 atoms with box lengths of 6.1698 (twice) and 7.0053; the other parameters are the
same as above, with 256 A and 64 B atoms. 
Configurations were saved from five MD runs of $10^6$ steps each, with $10^5$ steps
of equilibration, as for the smaller system. 
The solid is superheated, and only escapes from the crystal in the highest 
energy run on this time scale (Table  \ref{tab:MD}). The fraction of NSP
located in $|\nabla E|^2$ minimizations increases systematically from zero at the lowest energy,
where each minimization finds the crystal. 
The $I$ values associated with these runs are all much lower than
for the smaller supercell, indicating that the linear rise in $I$ above a threshold temperature
that has been observed for supercooled liquids is not simply a universal effect of temperature, 
but is specific to that region of configuration space. 
However, stationary points of index up to four are located in run four, where the 
system still does not escape from a single permutational isomer of
the crystal. On minimizing the energy from all the $|\nabla E|^2$
stationary points located in the latter run 97.8\%\ relax to the crystal, but nine other minima are also 
found. This result simply reflects the fact that minimizing $|\nabla E|^2$ can raise the energy sufficiently
for the configuration to escape from the crystal, even though the system is trapped there on a long
time scale.

In view of the above results, and the dominance of NSP's on minimizing $|\nabla E|^2$, we question whether
the dynamics of a supercooled liquid can usefully be described in terms of the basins of attraction
of $|\nabla E|^2$. 
Most of these basins do not correspond to stationary points of $E$ at all for the present system. 
It is also interesting to note that a more recent study instead used a `Newton' 
method to locate stationary points of any index.\cite{Grigera01}
This approach can potentially avoid the NSP problem, however few details were provided 
and the nature of this new mapping should also be carefully examined.

Furthermore, the configuration does not appear to be closer 
(in terms of distance) 
to stationary points of any particular index as the temperature varies.
Although our results do not support the claim 
that $I$ vanishes at a well-defined temperature above the glass transition, they do confirm that there is
a dramatic decrease in the number of different local minima sampled around this point. 
Since the prediction that $I$ should vanish at $T_{\rm MCT}$ is a mean-field result,\cite{Cavagna98}
it is perhaps not surprising that it is not precisely obeyed.
Similarly, the relaxation time scale for structural glasses does not 
actually diverge at $T_{\rm MCT}$ because activated processes can still occur 
below this temperature.
Of course, it should be remembered that all our low temperature results for 
the 256-atom cell correspond to non-equilibrium data, since there is a 
crystalline phase available.\cite{Middleton01b}

\section{Lennard-Jones Clusters}
\label{sect:clusters}

From simulations one can obtain the probability distributions of the system being in the 
basin of attraction of a minimum of energy $E$ at a temperature $T$. 
As one can evaluate the partition function of a minimum within the harmonic approximation, or
more accurately using anharmonic expressions,\cite{Doye95a,Calvo01e} the probability distributions
can be inverted to obtain distributions for the number of minima.\cite{Doye95a,Sciortino99a,Buchner99a} 

However, even if we could find a way of dividing up configuration space into 
basins around saddle points (i.e.\ without the problem of NSP's associated with the $|\nabla E|^2$ mapping),
we could not find the actual distributions of the number of saddles from simulation. 
The missing ingredient is
an expression for the partition function associated with the basin around 
a saddle,\cite{Zts} without which probability distributions of 
saddles obtained from simulation cannot be inverted.

Therefore, an alternative approach is needed to obtain distributions of saddle points.
Here we aim to obtain (near) complete sets of saddle points for a model finite system,
namely small Lennard-Jones clusters.
This task has been previously attempted for minima and transition states up to $N$=13 by Tsai and Jordan;\cite{Tsai93a}
since then larger databases have been obtained for some of these clusters.\cite{WalesD97b,Ball99a,Chekmarev01}
Although there are standard methods available to find minima and transition states, 
finding saddle points of a particular index represents a new challenge.
Eigenvector-following provides an efficient way to locate transition states, where we
search uphill in one direction while minimizing in the tangent space.\cite{Cerjan}
We have simply extended this approach to find a saddle point of index $I$ by searching uphill
in $I$ orthogonal directions, while minimizing in all other directions. All the uphill directions
were treated in the same way as for a transition state search, and were orthogonalized to the search
direction and gradient in the tangent space minimization, requiring only minor modifications of
our usual approach.\cite{WalesDMMW00}

\begin{table*}
\caption{\label{table:nsp} Number of saddle points of each index for LJ$_N$ clusters.
The numbers in italics are likely to be far from complete.}
\begin{ruledtabular}
\begin{tabular}{ccccccccccccccccc}
 & \multicolumn{16}{c}{stationary point index } \\
\cline{2-17} 
 $N$ &  0 &     1 &     2 &     3 &     4 &     5 &     6 &     7 &     8 &    9 &   10 &   11 &  12 & 13 & 14 & 15 \\
\hline
 4 &    1 &     1 &     2 &     1 &     1 &     0 &     0 \\
 5 &    1 &     2 &     4 &     6 &     6 &     2 &     1 &     0 \\
 6 &    2 &     3 &    13 &    24 &    30 &    26 &    16 &     5 &     1 &    0 \\
 7 &    4 &    12 &    44 &    98 &   168 &   190 &   168 &   101 &    45 &   11 &    1 &    0  \\
 8 &    8 &    42 &   179 &   494 &  1000 &  1458 &  1619 &  1334 &   852 &  388 &  125 &   26 &   1 &  0\\
 9 &   21 &   165 &   867 &  2820 &  6729 & 12093 & 16292 & 16578 & 13226 & 8286 & 4053 & 1444 & 376 & 56 &  1 & 0 \\
10 &   64 &   635 &  4074 & 16407 & 46277 & {\it 97183} \\
11 &  170 &  2424 & 17109 & {\it 47068} & \\
12 &  515 &  8607 & {\it 27957} \\
13 & 1509 & 28756 & {\it 88079} & & \\
\end{tabular}
\end{ruledtabular}
\end{table*}

To generate the sets of stationary points, we begin by obtaining samples of 
minima and transition states as in previous applications.\cite{Tsai93a,Miller99a,WalesDMMW00} 
From these sets of stationary points we search for index two saddles after randomly perturbing the
coordinates of the minima and transition states. We typically perform twenty such searches
for each stationary point. We then iteratively repeat this procedure for higher-index 
stationary points, at each stage performing searches from all stationary points of lower index.
This procedure is terminated when no stationary points of a particular index are found.

The sets of stationary points obtained in this manner are typically incomplete, 
and  the incompleteness is larger for stationary points of lower index, for 
which fewer searches have been conducted.
To converge the sets a reverse procedure was performed.
Starting from the stationary points of highest index ($I_{\rm max}$) searches are performed 
for saddle points of index $I_{\rm max}-1$ following a random perturbation.
Typically five searches from each stationary point are enough to ensure convergence.
Searches are then performed for stationary points of 
index $I_{\rm max}-2$ from all those of higher index, and so on until the searches for minima are completed.
The importance of this reverse procedure is evident, for example, 
from the approximately 50\% increase in the number of LJ$_{13}$ transition states located 
when the searches from higher-index saddle points were performed. 

\begin{figure}
\begin{center}
\includegraphics[width=8.2cm]{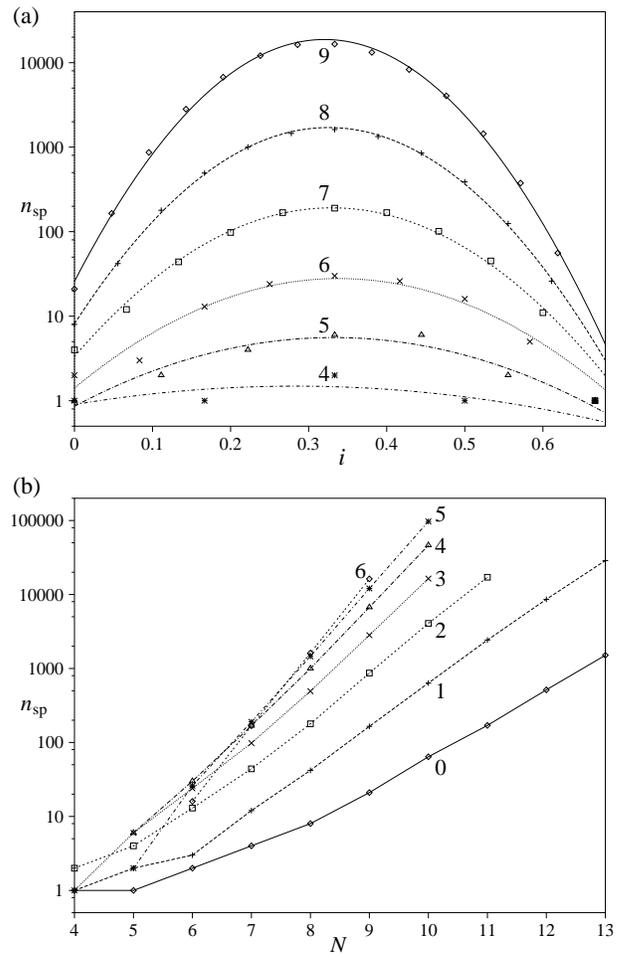}
\caption{\label{fig:nsp} 
The number of saddle points as a function of (a) the saddle point index and (b) the size.
In (a) the data points are from Table \ref{table:nsp}, the curves are Gaussian fits to the data, 
each curve is labelled by the cluster size and only sizes for which we have obtained 
complete distributions are represented. In (b) lines are labelled by the saddle point index.
}
\end{center}
\end{figure}

The numbers of stationary points as a function of the saddle point index are given 
in Table \ref{table:nsp}. 
We were able to find complete distributions of stationary points of any index 
for all clusters up to $N$=9. 
However, above this size the search had to be terminated at low index,
because the numbers of stationary points involved are too large
for a characterization of the whole distribution to be feasible.
For example, extrapolation suggests that for LJ$_{13}$ that there 
are of the order of $10^8$ saddle points of the most common index.
For these larger sizes the reverse procedure was still performed, but 
starting from the highest index for which we performed searches. 
As a result the number of saddle points of the highest index searched is likely to 
be much less than the true total, because no searches from saddle points of 
higher index have been performed. 

One particularly striking feature of the results is the large number of higher-index stationary 
points for systems of such small size, especially relative to the number of minima.
For example, LJ$_9$ has approximately 800 times more index seven saddles than minima.

To quantitatively probe these distributions we plot in Figure \ref{fig:nsp}a the number of saddle points
against the intensive measure of the saddle point index, $i$.
For a cluster, $i=I/(3N-6)$ because there are a further three zero 
eigenvalues corresponding to rotations.
As is clear from Figure \ref{fig:nsp}a the data fits very well to the Gaussian form:
\begin{equation}
\label{eq:Gaussian}
n_{\rm sp}(I)=n_{\rm sp}^{\rm max} \exp\left(-{\left(I-I_{\rm mid}\right)^2\over 2\sigma^2}\right).
\end{equation}
The agreement is less good for the smaller sizes, but this result is unsurprising considering the 
small number of stationary points.

The parameters of the Gaussian fits appear in Table \ref{table:Gaussian}. 
It is particularly noteworthy that the mid-points of the distributions 
in Figure \ref{fig:nsp}a are approximately constant. 
$i_{\rm mid}\approx 1/3$ which implies $I_{\rm mid}\approx N-2$.
Of course, the tail of the Gaussian is cut off at $I$=0, but it 
is also cut off beyond $2 I_{\rm mid}$. 
There are no stationary points for $I>I_{\rm max}=2N-4$.

Another interesting feature of the Gaussian distribution is that the standard deviation
of the distributions is only a weak function of $N$, scaling sublinearly with size.
Therefore, the distributions when considered as a function of $i$ (rather than $I$), 
as in Figure \ref{fig:nsp}, become narrower as the size increases.

\begin{table}
\caption{\label{table:Gaussian} Parameters for the Gaussian fits of the distributions $n_{\rm sp}(I)$.
Results are only included for sizes that have a complete distribution.}
\begin{ruledtabular}
\begin{tabular}{ccccc}
 $N$ & $n_{\rm sp}^{\rm max}$ & $I_{\rm mid}$ & $i_{\rm mid}$ & $\sigma$ \\
\hline
 4 &  1.49 & 1.700 & 0.283 & 1.699 \\
 5 &  5.59 & 2.989 & 0.332 & 1.545 \\
 6 &  27.9 & 4.063 & 0.339 & 1.663 \\
 7 & 191.4 & 4.944 & 0.330 & 1.738 \\
 8 &  1706 & 5.858 & 0.325 & 1.791 \\
 9 & 18782 & 6.728 & 0.320 & 1.853 \\
\end{tabular}
\end{ruledtabular}
\end{table}

Eq.\ (\ref{eq:Gaussian}) allows us to predict the ratio of the number of transition states to minima:
\begin{equation}
{n_{\rm ts}\over n_{\rm min}}=\exp\left({2I_{\rm mid}-1\over 2\sigma^2}\right)
                      \approx \exp\left({2N-5\over 2\sigma^2}\right).
\end{equation}
This ratio scales less than exponentially with $N$, because $\sigma$ is a weakly increasing function of $N$. 
The above equation can be rearranged to obtain an expression for $\sigma$:
\begin{equation}
\sigma\approx \sqrt{{N-5/2\over \log\left(n_{\rm ts}/n_{\rm min}\right)}}.
\end{equation}
The value of $\sigma$ thus obtained will, of course, involve a greater error than that obtained from 
fitting to the complete saddle point distribution, but it can be applied to the larger clusters for which 
we do not have complete distributions.
Again we find that $\sigma$ continues to increase slowly with $N$.

In Figure \ref{fig:nsp}b we show how the number of stationary points with a particular index
depends upon $N$. Most of the plots seem to tend to straight lines for larger $N$. 
Of course, there are likely to be significant deviations when $n_{\rm sp}(I)<100$, where the values
could reflect peculiarities of these small sizes.
Furthermore, there is insufficient data to definitely confirm an exponential scaling with size.

\begin{figure}
\begin{center}
\includegraphics[width=8.2cm]{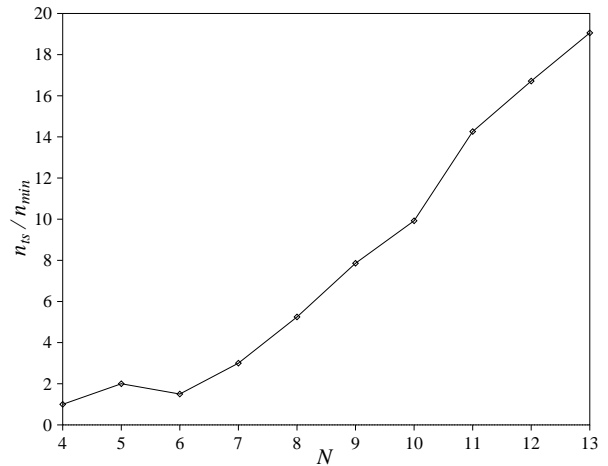}
\caption{\label{fig:ntsmin} The ratio of the number of minima to transition states as
a function of $N$.
}
\end{center}
\end{figure}

There is a theoretical expectation for $n_{\rm min}$ to scale exponentially with size.\cite{StillW82,Still99}
Here we present the simple argument of Ref.\ \onlinecite{StillW82}, which 
applies to a sufficiently large system.
If a system of $m N$ atoms can be divided into $m$ equivalent sub-systems of $N$ atoms, and the stable 
arrangements of the subsystem are effectively independent, then 
\begin{equation}
n_{\rm min}(m N)=n_{\rm min}(N)^m
\end{equation}
The solution of this equation is 
\begin{equation}
n_{\rm min}(N)=\exp(\alpha N).
\end{equation}
A similar argument can be given for the number of transition states.
Assuming the rearrangements associated with the transition states can be localized 
to one sub-cell, the whole $mN$-atom system will be at a transition state 
when one of the subsystems is at a transition state and the rest are at a minimum.
Therefore,
\begin{equation}
n_{\rm ts}(m N)=m\,n_{\rm min}(N)^{m-1} n_{\rm ts}(N) 
\end{equation}
The solution of this equation is
\begin{equation}
n_{\rm ts}(N)=N \exp(\alpha N).
\end{equation}
Hence, $n_{\rm ts}/n_{\rm min}$ grows linearly with size.
Our databases are consistent with this trend (Figure \ref{fig:ntsmin}).
The value of $\alpha$ is system dependent, and we illustrate this fact for the
13-atom cluster bound by the Morse potential, M$_{13}$, as a function of the 
range parameter, $\rho$. The results in Table \ref{tab:M13} supersede those in
previous work,\cite{WalesDMMW00} and were obtained using transition state searches
based on starting points obtained by considering hard sphere collisions, as described
elsewhere.\cite{Middleton01}

\begin{table}
\caption{Numbers of minima, $n_{\rm min}$, and transition states, $n_{\rm ts}$,
for M$_{13}$ clusters at three values of the range parameter $\rho$.}
\begin{center}
\begin{ruledtabular}
\begin{tabular}{crrrr}
$\rho$                                     & 3 \hfill     & 4  \hfill        & 6  \hfill           & 10    \hfill \\
\hline
$n_{\rm min}$                              & $7$   & $159$     & $1\,477$     & $10\,814$    \\
$n_{\rm ts}$                               & $47$  & $1\,366$  & $26\,431$    & $> 746\,283$ \\
$n_{\rm ts}/n_{\rm min}$                   & $6.7$ & $8.6$     & $17.9$       & $> 69.0$     \\
\end{tabular}
\end{ruledtabular}
\label{tab:M13}
\end{center}
\end{table}

This approach to finding expressions for the distribution of saddles
becomes more problematic for saddle points of higher index. 
As $I$ increases, the equations are increasingly hard to solve, 
as more combinations of saddle points of different index have to be considered,
and the assumption of sub-system independence becomes less plausible.
Therefore we need a different approach if we are to justify the 
Gaussian distribution for $n_{\rm sp}(I)$.

For any cluster there will always be a single stationary point of 
index $2N-4$ corresponding to a linear chain. 
For this configuration there are five zero Hessian eigenvalues 
(three translations and only two rotations), $N-1$ positive eigenvalues corresponding 
to bond stretches, as well as the $2N-4$ negative eigenvalues corresponding to bond-angle deformations.

The chain contains the minimum possible number of nearest-neighbour contacts 
for a bound configuration.
To produce any more negative eigenvalues would require the dissociation of an atom, 
but then the system would no longer correspond to a cluster of $N$ atoms.
Therefore, the linear chain must correspond to the saddle point with the highest index,
in agreement with the value of $I_{\rm max}$ that we found for each distribution.
In fact the linear chain is rather exceptional because of its five zero eigenvalues.
Any other configuration with $N-1$ bonds must be non-linear and 
so has $2N-5$ negative Hessian eigenvalues, six zeros, 
as well as the $N-1$ positive eigenvalues.
The linear chain's five zero eigenvalues allow the possibility of 
another negative eigenvalue.

\begin{figure}
\begin{center}
\includegraphics[width=8.2cm]{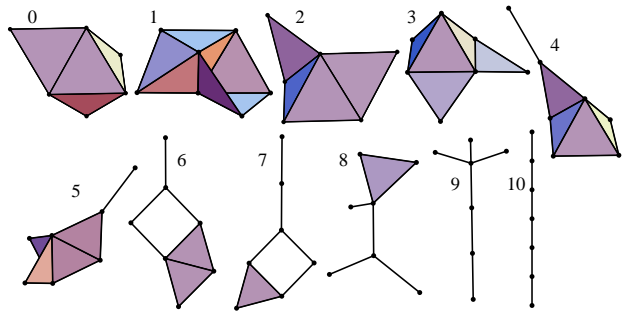}
\caption{\label{fig:spstructures} Some LJ$_7$ stationary points.
Each structure is labelled by the value of $I$, the index of the stationary point, and
each is at the midpoint of the energy distribution for stationary points of that index.}
\end{center}
\end{figure}

The analysis of the linear configuration suggests that the stationary point index 
corresponds to the number of bond-angle degrees of freedom that have negative eigenvalues.
The stationary points in Figure \ref{fig:spstructures} illustrate this trend.
Of course, there are some stationary points
where there are negative eigenvalues in the bond-stretch degrees of freedom, 
but these are in the minority.

This upper limit to $I$ also suggests that the Gaussian distribution may be a 
result of the number of different ways of choosing negative eigenvalues 
for $2N-5$ bond-angle degrees of freedom (assuming six zero eigenvalues and $N-1$ 
bond stretches). 
This assumption gives a binomial distribution 
\begin{equation}
n_{\rm sp}(I)={(2N-5)!\over (2N-5-I)! I!},
\end{equation}
which can be well approximated by a Gaussian with 
\begin{eqnarray}
I_{\rm mid}&=&N-5/2,\nonumber \\
n_{\rm sp}^{\rm max}&=&{(2N-5)!\over\left[(N-5/2)!\right]^2},\nonumber \\
\sigma&=&\sqrt{{N-5/2}\over 2}.
\end{eqnarray}
This simple analysis gives properties that are in good agreement
with our actual distributions. $I_{\rm mid}$ almost exactly matches
the observed value and $\sigma$ is a weakly increasing function of $N$.
The main error is the result that $n_{\rm sp}(0)=n_{\rm sp}(2N-5)=1$.
This error occurs because even when we have the maximum or minimum number of bond-angle degrees
of freedom with negative eigenvalues, there are still a number of different structural
ways that this can be achieved (roughly $\exp(\alpha N)$ in fact). 
Therefore, to correct for this error, the above expression for $n_{\rm sp}^{\rm max}$ 
can be multiplied by an exponential function fitted to the number of minima.
Also, the Gaussian fit to the binomial distribution gives the erroneous prediction 
that $n_{\rm ts}/n_{\rm min}$ is independent of $N$.
However, the binomial distribution itself gives the correct result, while
the Gaussian approximation begins to break down at the tails of the distribution.

\begin{figure}
\begin{center}
\includegraphics[width=8.2cm]{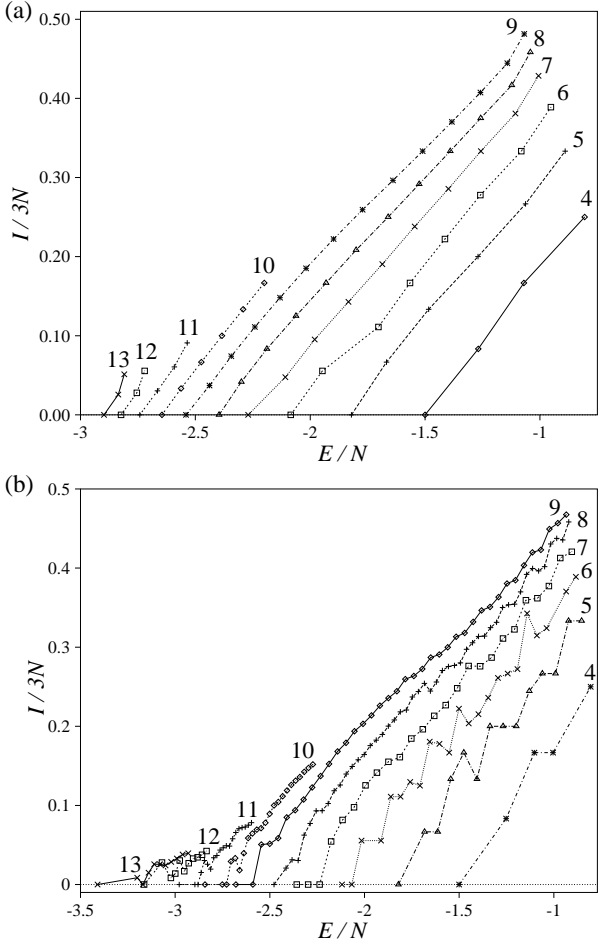}
\caption{\label{fig:IvE} $I/3N$ versus $E/N$.
In (a) the points are averages over $I$ and in
(b) averages over the energy. 
In (b) the curves are cut off at $I_{\rm max}-1/2$ 
for those sizes that have incomplete saddle point distributions.
}
\end{center}
\end{figure}

One quantity that has been focussed upon in previous studies of saddle points in glasses
has been the variation of the saddle point index with potential energy.\cite{Angelani00,Broderix00,Grigera01}
The averaging performed to obtain this function can be done in two ways. 
One can either look at $\langle E(I)\rangle$, the average energy of saddles with index $I$ (Figure \ref{fig:IvE}a), 
or at $\langle I(E)\rangle$, the average index of stationary points with energy $E$ (Figure \ref{fig:IvE}b).
For $N\ge 10$ as the energy increases the latter function saturates at the highest index 
for which we have performed searches.
To avoid this effect of the incomplete distributions, we have cut off the function at $I_{\rm max}-1/2$.

\begin{figure}
\begin{center}
\includegraphics[width=8.2cm]{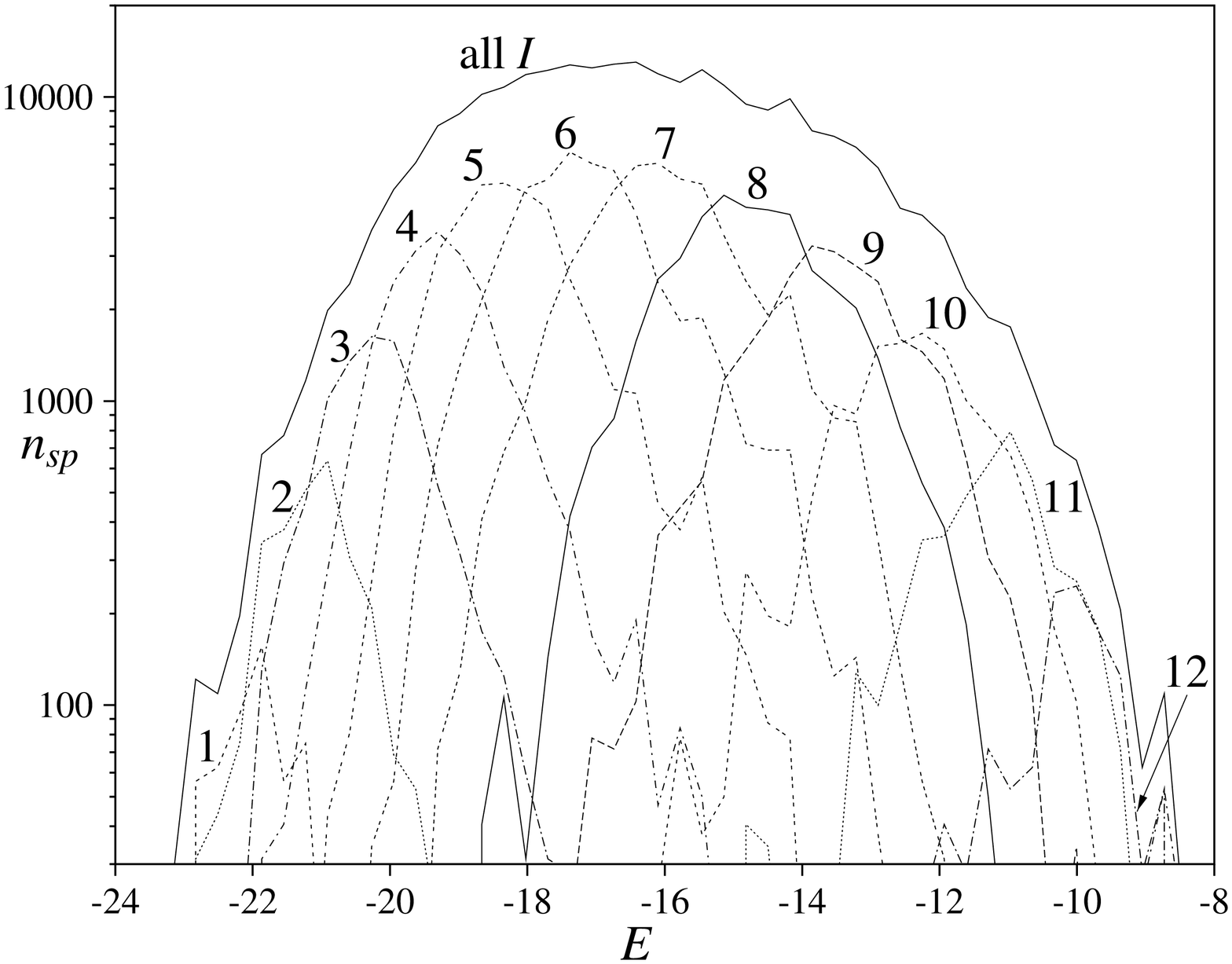}
\caption{\label{fig:nspvE} 
The number of LJ$_9$ saddle points of index $I$ as a function of energy.
Each curve is labelled by the index $I$, except for the sum of all the distributions.}
\end{center}
\end{figure}

Both averages show an approximately linear increase of the energy with saddle point index,
similar to that observed for supercooled liquids.\cite{Angelani00,Broderix00,Grigera01} 
However, the slope is somewhat lower for $\langle I(E)\rangle$, 
because the energy at which saddle points of a particular index are the most
common does not necessarily correspond to the energy at 
the maximum of their distribution, i.e.\ $\langle E(I)\rangle$ (Figure \ref{fig:nspvE}).
For $I<I_{\rm mid}$ the majority of the range for which they are most numerous has 
$E<\langle E(I)\rangle$ and for $I>I_{\rm mid}$ has $E>\langle E(I)\rangle$.

The slopes of curves such as those in Figure \ref{fig:IvE} have been interpreted in 
terms of a single characteristic barrier height for the PES.\cite{Broderix00,Angelani01,Grigera01} 
As the lines in Figure \ref{fig:IvE} have similar slopes this `barrier' is only a weak function of cluster size.
However, there is of course a distribution of barrier heights.\cite{Middleton01,Mousseau00}
Furthermore, the experimentally-observed activation energy to structural relaxation is 
unlikely to correspond to a barrier associated with a single rearrangement, 
but to the overall barrier associated with a sequence of rearrangements.\cite{Miller99b,WalesDMMW00}
It is also easy to show that the average barrier between minima and transition
states, $\langle \Delta \rangle$, is different from the slope of Figure \ref{fig:IvE}a:
\begin{eqnarray}
\label{eq:barrier}
\langle \Delta\rangle&=&\langle E_{\rm ts}\rangle - \sum_i n^i_{\rm ts}E^i_{\rm min}/2n_{\rm ts}\\
&\not=&\langle E_{\rm ts}\rangle - \langle E_{\rm min}\rangle,
\end{eqnarray}
where $E^i_{\rm min}$ is the energy of minimum $i$ and $n_{\rm ts}^i$ is the number of transition
states connected to that minimum.
$\langle \Delta\rangle$ is likely to be larger because the average over minima in the second
term of equation (\ref{eq:barrier}) is usually weighted towards the lower-energy minima,
since they are connected to more transition states.
For example, for LJ$_{13}$ $\langle \Delta\rangle$=$1.771\epsilon$, whereas 
$\langle E_{\rm ts}\rangle - \langle E_{\rm min}\rangle$=$0.728\epsilon$. 

\begin{figure}
\begin{center}
\includegraphics[width=8.2cm]{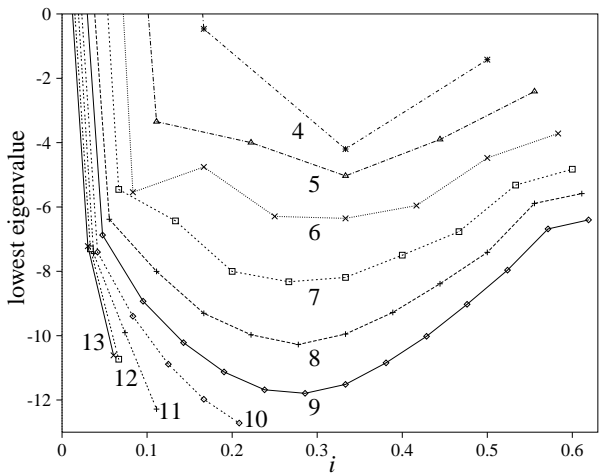}
\caption{\label{fig:lambda} 
The average of the lowest eigenvalue of the Hessian for saddle points of the same index
as a function of $i$.
The different curves correspond to different sizes as labelled.
The eigenvalues are in units of $\epsilon/\sigma^2$.}
\end{center}
\end{figure}

Another property of saddle points that has received attention is the lowest eigenvalue of the Hessian. 
For a binary Lennard-Jones liquid a linear decrease in the average value of the lowest Hessian eigenvalue was observed as the 
energy is increased above the threshold energy at which higher-index saddle points were first
observed.\cite{Broderix00} This result is equivalent to a linear decrease with $i$.
The behaviour of this property for our cluster saddle points is depicted in Figure \ref{fig:lambda}.
The curves seem to have a common form, in which the lowest eigenvalue reaches
a minimum close to $I_{\rm mid}$. However, these values of $i$ are much larger than those
probed in Ref. \onlinecite{Broderix00}, and the initial parts of the curves in Figure \ref{fig:lambda}
seem to show greater linearity as the size is increased.

\section{Discussion}
\label{sect:discuss}

So far we have examined the properties of a particular mapping that attempts to 
provide a definition of the neighbourhood of a saddle point, and looked at the properties
of higher-index saddle points for systems where we can obtain complete distributions. 
Here, we want to think more about the role of higher-index saddle points in the dynamics,
assuming that one could find a mapping that divides all of the PES in neighbourhoods around the closest saddle.
First we review the relevant aspects of the inherent structure approach to the dynamics,
i.e.\ in terms of the dynamics of transitions between the basins of attraction surrounding minima.
This approach can be formulated in term of a master equation\cite{vanK} 
describing the evolution of the occupation probability of a particular minimum 
in terms of the rates of probability flow into and out of that minimum:\cite{BerryK95,WalesDMMW00}
\begin{equation}
\label{eq:ME}
{dP_i(t)\over dt}=\sum_{j\ne i}\left[k_{ij}P_j(t)-k_{ji}P_i(t)\right],
\end{equation}
where $P_i$ is the occupation probability of the basin around minimum $i$ 
and $k_{ij}$ is the rate constant for a transition from basin $j$ to basin $i$.
This set of equations can then be solved to give a complete picture of the inter-minimum dynamics.

The only assumption that this approach relies upon is the Markovian nature of the underlying dynamics,
i.e.\ the probability of the transition $i\to j$ must not depend on the history of 
reaching minimum $i$, so that the $k_{ij}$ are constants for a given temperature or energy. 
This assumption will certainly be true when states within a basin of attraction
equilibrate on a time scale faster than transitions to different minima, 
i.e.\ $\tau_{\rm inter}\gg\tau_{\rm intra}$, where these two time scales 
are for interbasin and intrabasin relaxation.
Indeed, it is this separation of time scales that makes the inherent structure approach
a natural way to describe the dynamics.
The breakdown of the Markovian character of the interbasin dynamics 
places an upper bound on the temperature for which this approach is applicable.
Previous results for small alkali halide and LJ clusters show reasonable agreement
between MD rates and model rate theory for relatively high energies.\cite{Rose92,Miller97}

In the above theory for isomerization rates there is no requirement for the system to
lie close to the minimum in configuration space. During the occupancy of a given catchment
basin the system could be found, on average, close to the boundary. 
Therefore, the increase in $I$ with temperature seen for supercooled liquids above a threshold
temperature does not necessarily imply that the inherent structure dynamics approach has broken down.
Rather the test is whether the interbasin dynamics are no longer Markovian.

One of the main advantages of equation (\ref{eq:ME}) is that we can draw on the mature field
of unimolecular rate theory\cite{Baer96} to calculate the $k_{ij}$.
For example, the classical limits for the microcanonical and canonical
Rice-Ramsperger-Kassel-Marcus rate constants in the harmonic approximation are\cite{Baer96}
\begin{eqnarray}
\label{eq:RRKM}
k(E) &=& \left( E-E^\dag \over E \right)^{s-1} { \bar{\nu}^s \over \bar{\nu}^{s-1}_\dag }, \nonumber \\
k(T) &=& \exp( -E^\dag/kT ) { \bar{\nu}^s \over \bar{\nu}^{s-1}_\dag }, 
\end{eqnarray}
where $E$ and $E^\dag$ are the total energy and the potential energy of the transition state relative to the energy
of the minimum, $s$ is the number of vibrational degrees of freedom, and $\bar{\nu}$ and $\bar{\nu}_\dag$ are the geometric
mean vibrational frequencies of the minimum and transition state, respectively.

Angelani {\it et al.\/} have argued that 
``diffusion is entropy driven, even below $T_{\rm MCT}$'' based on the observation
that the instantaneous potential energy of the system lies 
well above the SP's obtained by minimizing $|\nabla E|^2$.\cite{Angelani00}
However, increasingly large values of the total energy are needed
to obtain significant rates as the number of degrees of freedom increases (see for example \cite{Baba97}),
as is clear from equations (\ref{eq:RRKM}).
To maintain a finite microcanonical rate constant
as $s$ rises the ratio $E/E^\dag$ must increase.
From the canonical viewpoint, the higher total energy is simply required to maintain a constant temperature
as the system size grows.

In the microcanonical ensemble the entropy is the appropriate thermodynamic potential, while
in the canonical ensemble we must instead consider free energy barriers,
although it should be remembered that the two ensembles are equivalent in the bulk limit. 
The relative energies of the transition state and minimum, and the widths of the minimum and transition state valley 
contained in the ratio of frequencies, contribute to the rate constant for an elementary unimolecular
isomerization in both ensembles. Both are included in the standard rate expressions above.
However, the relative importance of different
contributions to $k(E)$ cannot be assessed without considering how the different terms 
scale with system size.
For the canonical rate constant $k(T)$, where it is meaningful to consider separate entropic and energetic
contributions to the free energy barrier, the total energy does not enter.
The energy difference, $E^\dag$, for a particular class of rearrangement should be an intensive
rather than an extensive quantity.

For normal liquids above the melting point Dzugutov has demonstrated a strong empirical
correlation between the entropy and the diffusion constant.\cite{Dzugutov96}
Other recent simulation studies\cite{Scala00,Sastry00a,Sastry01}
have been interpreted using the Adam-Gibbs model,\cite{AdamG65}
where the entropy, or part of it, enters in a rather different way through heuristic arguments.
On the other hand, atomic diffusion in solids is routinely treated by Vineyard's approach,\cite{Vineyard57}
which is simply an application of conventional unimolecular rate theory to bulk.
There is clearly a pressing need to determine the limit of applicability of standard
rate theory to supercooled liquids, and to develop alternative approaches with a firm
microscopic basis where necessary.

Cavagna has suggested that when $\tau_{\rm inter}\sim\tau_{\rm intra}$ it is more appropriate to
think about the dynamics in terms of transitions between the neighbourhoods of saddle points.\cite{Cavagna01}
As equation (\ref{eq:ME}) can be applied to any partition of the PES into `basins', 
not just those surrounding minima, a similar formalism can be developed as long as the
inter-saddle dynamics are Markovian. However, therein lies the problem. 
If the residence times
in the basins of attraction surrounding a minimum are too short for equilibrium 
between the vibrational modes to be established, it seems even less likely that the
necessary separation of time scales will hold for basins
surrounding saddle points. 
By definition, there are forces acting to take the system out of such regions.
Furthermore, in contrast to the case of isomerizations between minima, there is no established theory
for transition rates between saddles. In fact, it is unclear how to calculate the partition
function for the catchment basin of a saddle, which would surely be necessary to evaluate
rate constants.
It is therefore hard to see how this view of the dynamics can 
be put on a quantitative footing.
In his contribution Cavagna simply speculated that the 
intersaddle rate constants would increase as $I$ increased.\cite{Cavagna01}

One other possible criticism of the `saddles-ruled' approach of Cavagna is that it
seems to ignore many of the effects of the topography of the PES, and so conflicts with
much of the recent work that has emphasised the importance of this
topography.\cite{Sastry98,Debenedetti01,Sastry01} 
For example, Cavagna suggests that the origin of strong and fragile 
behaviour\cite{Angell91} 
simply lies in the value of the thermal energy at the temperature for which saddles begin
to be frequently sampled relative to the ``barrier'' obtained from the slope of the line
giving the averaged dependence of the index upon the saddle point energy.\cite{Cavagna01} 

It is also too simplistic
to suggest that the index of a saddle indicates the `number of diffusive directions'.\cite{Angelani00}
In fact, rearrangements may be both diffusive and non-diffusive in nature,\cite{Middleton01}
and the character can only be diagnosed by calculating steepest-descent paths
connecting the relevant stationary points, not from purely local information. 
The non-diffusive transitions typically involve an atom moving slightly within the cage 
of its neighbours.

\section{Conclusions}

We have examined in detail the behaviour of the $|\nabla E|^2$ mapping of 
configurations to saddle points, and find that it has a number of shortcomings.
We agree with Cavagna\cite{Cavagna01} that the mapping cannot partition the whole of the PES into
basins surrounding the saddle points, as claimed by Angelani {\it et al.\/}.\cite{Angelani00} 
In fact, for the systems we have studied
the vast majority of configuration space sampled by the supercooled
liquid is mapped on to points that are not stationary points of the PES. 
Furthermore, the saddle points obtained by this mapping are no closer to the initial 
configuration than are the nearest transition state or minimum.
Therefore, the mapping does not seem to satisfy the requirement of dividing the PES into
neighbourhoods around the closest saddle.
We have also pointed out problems with a number of interpretations suggested in previous work.

A more straightforward way to characterize the properties of higher-index saddles is
for systems where complete distributions of saddle points can be obtained. Our results
for LJ clusters reveal that the distributions are a Gaussian function of 
the index, as well as of the energy.\cite{Sciortino99a,Buchner99a} 
Gaussian index distributions have also been found for random matrices.\cite{Cavagna00}
We have suggested an explanation for this distribution in terms of the number
of possible ways of assigning negative eigenvalues to a set of bond-angle degrees of freedom. 
We find an approximately linear relationship between the potential
energy of the saddle point and its index, similar to the results for supercooled liquids.

We do not think that a description of the dynamics in terms of inter-saddle, 
rather than inter-minimum, transitions will offer any advantages, 
even if a proper partitioning of the PES into catchment basins of saddles can be devised.
Although the temperature at which the dynamics become non-Markovian, 
and hence the linear master equation formalism breaks down, has not yet been established 
it may well be at temperatures where the 
time scale for transitions between minima is comparable to the time for
equilibrium to be established between the vibrational modes. 
Below this regime we should be able to apply 
standard unimolecular rate theory, but above it any approach focussing on uncorrelated transitions
between local regions of configuration space is likely to fail.
Other approaches, such as mode-coupling theory,\cite{Gotze91,Kob97,Gotze99} will then be more appropriate.

\acknowledgements
JPKD is grateful to Emmanuel College, Cambridge and the Royal Society for financial support.
The authors would like to Francesco Sciortino and Juan Garrahan for useful discussions.


\begin{thebibliography}{10}
\bibitem{DJWwhoops}
The stationary points of the M\"{u}ller-Brown surface reported in Ref.\
  \onlinecite{Wales94b} are correct but their energies are not. This bug does
  not affect any conclusions of Ref.\ \onlinecite{Wales94b}.
\bibitem{Zts}
The contribution to the partition function for a basin surrounding a saddle
  point from the space orthogonal to the unstable modes is easy to evaluate,
  and the expression is the same as for a minimum with the dimension of this
  reduced space. However, the contribution from the space spanned by the
  unstable modes is unclear. One approximation that has been used for
  transition states is to assume a flat valley of a certain length in the
  direction of the transition vector,\cite{Chekmarev93,Doye95a} and this
  approach could be extended to all unstable modes.
\end{thebibliography}
\end{document}